# The Planck law for particle with rest mass

Piero Chiarelli

*National Council of Research of Italy, Area of Pisa, 56124 Pisa, Moruzzi 1, Italy*

*Interdepartmental Center "E.Piaggio" University of Pisa*
Phone: +39-050-315-2359
Fax: +39-050-315-2166

Email: pchiare@ifc.cnr.it.

Abstract: In this work the author derives the Plank law for particles with rest mass that in the limit of photons with null rest mass leads to the known Plank law for the black body radiation. The paper shows that the relation has a pure quantum origin deriving it from the quantum potential of the hydrodynamic model. The paper puts under light how these laws originates by the scale invariance breaking due to the quantum potential. The paper also briefly analyzes the consequences of this symmetry breaking on the small-size black holes formation.

**PACS**: 42.50.Lc, 04.60.-m

**Keywords:** Plank law**,** quantum stochastic hydrodynamic model, vacuum mass fluctuations, minimum black hole mass, black hole quantization

## 1. Introduction

The emerging of the quantum mechanics in the classical macroscopic world had one of the most important confirmations by the Planck's law of the black body radiation that dates back to the first year of the last century [1]. Despite the great success of the quantum mechanics, the theoretical connection between it and the classical one, still is underway [2-4].
One current of thought is represented by the "deterministic" approach to quantum mechanics that analyzes how the quantum equations are the generalization of the classical one [4-13] where the non-locality is introduced in various ways, the Madelung quantum potential [4, 6-7], the Nelson's osmotic potential, the Bohm-Hylei quantum potential or the Paris and Wu fifth-time parameter.
A great help in explaining the origin of the quantum mechanics comes from the quantum hydrodynamic analogy (QHA) [4, 6-7] that shows how the non-local restrictions come in playing from the quantization of vortices [4] and by the elastic-like energy arising by the quantum pseudo-potential (QP). The study of the quantum non-locality brought by the QP in the QHA makes more clear the connection between quantum concepts (probabilities) and classical ones (e.g., trajectories) [8] .
The QHA and similar theories find their validity in the fact that they help in explaining quantum phenomena that cannot be easily described by the usual formalism. They are multiple tunneling [19], critical phenomena at zero temperature [20], mesoscopic physics [21-22], quantum dispersive phenomena in semiconductors [24], quantum field theoretical regularization procedure [23] and the quantization of Gauge fields, without gauge fixing and without ensuing the Faddeev-Popov ghost [25].
The objective of the work is to show the deeper level of connection between the quantum characteristics of the vacuum and the Planck law allowing the theoretical derivation of it from a general law holding for particle with rest mass. The result put under light the scale invariance breaking of the QP and how it determines the quantum property of the vacuum submitted to thermal fluctuations.
This result is not without advantages: it allows the correct stochastic generalization of the QHA theory that is able to comprehend the classical mechanics as a large scale limiting dynamics and to develop a criterion able to introduce and explain the quantum hindering effect on the small-size black hole.
The work is organized as follows, in the first part the spectrum of the thermal fluctuations of vacuum wave function modulus (i.e., particle density (PD)) with mass $m$ at a temperature T is deduced by using the quantum potential of the QHA. Then, the result is generalized to the relativistic limit and applied to photons. Finally, it is shown that in the cased of a black hole of Planck mass, the QP energy is equal to its gravitational energy changed of sign and hence, that the total energy is null hindering its formation.

## 2. Quantum hydrodynamic analogy

In this section the hydrodynamic representation of quantum mechanic is briefly introduced in order to derive the Plank law for the vacuum thermal field of particle with rest mass *m*.

The QHA describing the motion of a particle density $n_{(q,t)} = |\cdot|^2_{(q,t)}$ with velocity $\dot{q} = \frac{\nabla S_{(q,t)}}{m}$, is equivalent to the Schrödinger equation applied to a wave function $\cdot_{(q,t)} = |\cdot|_{(q,t)} \, exp[\frac{i}{\hbar} S_{(q,t)}]$, and is defined by the equations [ 4]

$$\partial_t n_{(q,t)} + \nabla \bullet ( n_{(q,t)} \dot{q} ) = 0 , \quad (1)$$

$$\dot{q} = \frac{\partial H}{\partial p} = \frac{p}{m} = \frac{\nabla S_{(q,t)}}{m} , \quad (2)$$

$$\dot{p} = -\nabla ( H + V_{qu} ) , \quad (3)$$

where

$$H = \frac{p \bullet p}{2m} + V_{(q)} \quad (4)$$

is the Hamiltonian of the system and where $V_{qu}$ is the quantum pseudo-potential that reads

$$V_{qu} = -( \frac{\hbar^2}{2m} ) n^{-1/2} \nabla \bullet \nabla n^{1/2} . \quad (5)$$

If we consider the presence of thermal vacuum fluctuations of the PDF $n_{(q,t)} = |\cdot|^2_{(q,t)}$, equation (1) can be obtained as the deterministic limit of the stochastic equation [26]

$$\partial_t n_{(q,t)} = -\nabla \bullet ( n_{(q,t)} \dot{q} ) + ¦_{(q,t,T)} \quad (6)$$

One of the major obstacle in order to define the SPDE (12) is to define the noise term due to the thermal fluctuations induce by a thermostat on the vacuum. For the sufficiently general case, to be of practical interest, $¦_{(q,t,T)}$ can be assumed Gaussian with null correlation time and the noises, on different co-ordinates, independent. In this case, the stochastic partial differential conservation equation is completed by the relations [38]

$$< ¦_{(q_{\sim},t)}, ¦_{(q + ª, t+²)} > = < ¦_{(q_{\sim})}, ¦_{(q)} > G(ª) ¢ (²) ¢_{\sim} \quad (7)$$

$$S = \int_{t_0}^{t} dt ( \frac{p \bullet p}{2m} - V_{(q)} - V_{qu(n)} ) \quad (8)$$

where $T$ is the noise amplitude parameter (e.g., the temperature of an ideal gas thermostat put in equilibrium with the vacuum [26]) and $G(\lambda)$ is the shape of the spatial correlation function of $\eta$.

As already shown [26], due to the quantum potential, the shape of the spatial correlation function cannot be a delta-function (so that the spatial dependence of the noise $\eta$ cannot be white) but has the the correlation function

$$\lim_{T \to 0} G(\lambda) = exp[-(\frac{\lambda}{\lambda_c})^2] \qquad (9)$$

The noise spatial correlation function (9), is a direct consequence of the derivatives present into the quantum potential that give rise to an elastic-like contribution to the system energy, coming from the PD derivative, that reads [23]

$$\overline{H}_{qu} = \int_{-\infty}^{\infty} n_{(q,t)} V_{qu(q,t)} dq = -\int_{-\infty}^{\infty} n_{(q,t)}^{1/2} (\frac{\hbar^2}{2m}) \nabla \bullet \nabla n_{(q,t)}^{1/2} dq , \qquad (10)$$

where a large derivative of $n_{(q,t)}$ leads to high quantum potential energy. This can be easily checked by calculating the quantum potential of the pure sinusoidal mode (e.g., mono-dimensional case)

$$n = n_0 \, cos \frac{2\pi}{\lambda} q \qquad (11)$$

leading to

$$V_{qu} = -(\frac{\hbar^2}{2m}) \left( cos^2 \frac{2\pi}{\lambda} q \right)^{-1/2} \nabla \bullet \nabla \left( cos^2 \frac{2\pi}{\lambda} q \right)^{1/2} = \frac{\hbar^2}{2m} \left( \frac{2\pi}{\lambda} \right)^2 \qquad (12)$$

showing that the energy increases as the inverse squared of the the wave length of fluctuation.

Therefore, particle density fluctuations mode containing very high spatial frequencies (i.e., $\lambda \to 0$) can lead to a whatever large quantum potential energy even in the case of vanishing fluctuations amplitude (i.e., $T \to 0$).

In this case the convergence of equations (6) to the deterministic limit (1-3) (i.e., the standard quantum mechanics) may not happen since a diverging quantum potential energy for the zero limiting amplitude fluctuations can lead to a finite energy contribution to the system in the null noise limit.

In order to eliminating this unphysical possibility, the additional conditions (9) comes into the set of the relations leading to the stochastic generalization of quantum mechanics [26].

The demonstration brings a quite heavy stochastic calculations [26], but a more simple and straight justification of the correlation function (9) can come by considering the spectrum of the PD fluctuations of the vacuum. Since each, independent, component of spatial frequency $k = \frac{2\pi}{\lambda}$ brings a quantum potential energy contribution (12), the probability of happening is $p = \boldsymbol{exp}\left[-\frac{E}{kT}\right]$, that reads

$$p = exp\left[-\frac{E}{kT}\right] = exp\left[-\frac{< V_{(q)} + Vqu >}{kT}\right] \qquad (13)$$

leading in the empty vacuum (i.e., $V_{(q)} = 0$ ) to the expression:

$$p \propto exp\left[-\frac{<Vqu>}{kT}\right] = exp\left[-\frac{<\frac{\hbar^2}{2m}\left(\frac{2\text{®}}{\text{a}}\right)^2>}{kT}\right] \qquad (14)$$

$$= exp\left[-\frac{\hbar^2}{2mkT}\left(\frac{2\text{®}}{\text{a}}\right)^2\right] = exp\left[-\left(\frac{\text{®}\text{a}_c}{\text{a}}\right)^2\right] = exp\left[-\frac{\hbar}{2mc}\frac{\hbar c}{\text{a}kT}\right]$$

where

$$\text{a}_c = 2\frac{\hbar}{(2mkT)^{1/2}} \qquad (15)$$

From (14) it comes out that the spatial frequency spectrum $S(k) \propto p(\frac{2\text{®}}{\text{a}})$ is not white since fluctuations with smaller wave length have larger energy so that when $\text{a}$ is smaller that $\text{a}_c$ their amplitude go quickly to zero.

Given the spatial frequency spectrum $S(k) \propto p(\frac{2\text{®}}{\text{a}})$, the spatial noise correlation function of the vacuum (particle density) fluctuation reads

$$G_{(\text{a})} \propto \int_{-\infty}^{+\infty} exp[ik\text{a}]S_{(k)}dk \propto \int_{-\infty}^{+\infty} exp[ik\text{a}]\, exp\left[-\left(k\frac{\text{a}_c}{2}\right)^2\right]dk$$
$$\propto \frac{\text{®}^{1/2}}{\text{a}_c} exp\left[-\left(\frac{\text{a}}{\text{a}_c}\right)^2\right] \qquad (16)$$

that leads to (9 ).

If we want to find the limit of (14) for a photon we have to consider the relativistic case whose sinusoidal PDF fluctuation mode in 4-D space reads

$$\text{n}^{-1/2} = |\cdot\ | = |\cdot_0| cos(k_i q_i - \text{µ}t)$$

with $|k|v = \text{µ}$ that by a rotation we can reduce to

$$\text{n}^{-1/2} = |\cdot\ | = |\cdot_0| cos\left(\frac{2\text{®}}{\text{a}}(q - vt)\right)|. \qquad (17)$$

where $\frac{2\text{®}}{\text{a}} = |k| \equiv k$ ,nunited to the relativistic expression of the quantum energy

$$H = \pm\sqrt{m^2c^4 + p^2c^2 - \hbar^2c^2\frac{\partial_{\text{«}}\partial^{\text{«}}\text{n}^{1/2}}{\text{n}^{1/2}}}$$

$$H = m\text{¥}c^2\left(\sqrt{1 - \frac{\hbar^2}{(mc)^2}\frac{\partial_{\text{«}}\partial^{\text{«}}|\cdot\ |}{|\cdot\ |}}\right) = m\text{¥}c^2\left(\sqrt{1 + \frac{V_{qu}}{mc^2}}\right) \qquad (18)$$

where quantum potential $V_{qu}$ [27] reads

$$V_{qu} = -\frac{\hbar^2}{m} \text{n}^{-1/2} \left( \frac{1}{c^2} \frac{\partial^2}{\partial t^2} - \nabla \bullet \nabla \right) \text{n}^{1/2} = -\frac{\hbar^2}{m} \left( \left( \frac{2\text{®}}{\text{a}} \right)^2 \left( 1 - \frac{v^2}{c^2} \right) \right) \tag{19}$$

leads to the probability of the $\text{a}$ – fluctuation mode

$$p = exp\left[ -\frac{E}{kT} \right] = exp\left[ -\frac{m\text{¥}c^2 \left( \sqrt{1 + \frac{V_{qu}}{mc^2}} \right)}{kT} \right]$$

$$= exp\left[ -\frac{m\text{¥}c^2 \left( \sqrt{1 - \left( \frac{\hbar}{mc} \right)^2 \left( \frac{2\text{®}}{\text{a}} \right)^2} \right)}{kT} \right] \tag{20}$$

Thence, the probability $P$ of having a total energy $E$ (with $n$ particles) in a range of wave length between $\text{a}$ and $\text{a} + d\text{a}$ is

$$P_{(E(\text{a},n)} = exp\left[ -\frac{nm\text{¥}c^2 \left( \sqrt{1 - \left( \frac{\hbar}{mc} \right)^2 \left( \frac{2\text{®}}{\text{a}} \right)^2} \right)}{kT} \right] \tag{21}$$

that leads to the mean energy between $\text{a}$ and $\text{a} + d\text{a}$

$$< E >= \frac{m\text{¥}c^2 \left( \sqrt{1 - \left( \frac{\hbar}{mc} \right)^2 \left( \frac{2\text{®}}{\text{a}} \right)^2} \right)}{exp\left[ \frac{m\text{¥}c^2 \left( \sqrt{1 - \left( \frac{\hbar}{mc} \right)^2 \left( \frac{2\text{®}}{\text{a}} \right)^2} \right)}{kT} \right] - 1} \tag{22}$$

Moreover, calculating the density of modes with wave length $\text{a}$ or wave vector modulus $\text{k} = \frac{2\text{®}}{\text{a}}$ in the three dimensional space

$$\text{n}_{(\text{k})} = \frac{1}{V} \frac{dN_{(\text{k})}}{d\text{k}} = \frac{1}{2} \frac{\text{k}^2}{\text{®}^2} \tag{23}$$

it follows that the spectral density of the thermal PD (wave function modulus) fluctuation of the vacuum is

$$°_{(k)} = \mathrm{n}_{(k)} < E >= \frac{\frac{2m¥c^2}{a^2}\left(\sqrt{1-\left(\frac{\hbar}{mc}\right)^2\left(\frac{2®}{a}\right)^2}\right)}{exp\left[\frac{m¥c^2\left(\sqrt{1-\left(\frac{\hbar}{mc}\right)^2\left(\frac{2®}{a}\right)^2}\right)}{kT}\right]-1} \qquad (24)$$

Since in the case of a photon we have that

$$m = 0,$$

$$\frac{p}{c} = \frac{\hbar k}{c} = \frac{h}{a c}, \qquad (25)$$

$$\mathrm{n}^{-1/2} =| \cdot \ |=| \cdot\ _0 \,| cos\left(\frac{2®}{a}(q-ct)\right), \qquad (26)$$

$$V_{qu} = -\frac{\hbar^2}{m}\frac{\partial_{«}\partial^{«}\mathrm{n}^{1/2}}{\mathrm{n}^{1/2}} = -\frac{\hbar^2}{m}\mathrm{n}^{-1/2}\left(\frac{1}{c^2}\frac{\partial^2}{\partial t^2} - \nabla \bullet \nabla\right)\mathrm{n}^{1/2} = 0 \qquad (27)$$

$$H = \sqrt{m^2c^4 + p^2c^2 - \hbar^2c^2\frac{\partial_{«}\partial^{«}\mathrm{n}^{1/2}}{\mathrm{n}^{1/2}}} = \hbar µ, \qquad (28)$$

and thence

$$p \propto exp\left[-\frac{\hbar µ}{kT}\right] \qquad (29)$$

that leads to

$$< E >= \frac{\frac{\hbar c}{a}}{exp\left[\frac{\hbar c}{a kT}\right]-1} = \frac{\hbar µ}{exp\left[\frac{\hbar µ}{kT}\right]-1} \qquad (30)$$

and to

$$°_{(µ)} = 2\frac{°_{(k)}}{c} = 2\frac{\mathrm{n}_{(k)}}{c} < E >= \frac{µ^2}{®^2c^3}\frac{\hbar µ}{exp\left[\frac{\hbar µ}{kT}\right]-1}. \qquad (31)$$

where the factor 2 is due to the two polarization state of a photon, while the scalar wave function density has only one mode for each wave vector k .

## 3. The quantum scale invariance breaking and the minimum black hole mass

The quantum potential is clearly a "system quantity" that has strong physical reality. The fact that the vacuum fluctuations are discriminated on the foot of their wave lengths leads to the breaking of scale invariance of the classical physics. This practically implies that the large scale phenomena cannot be replicated on very small one. Applying this concept to one of the most meaningful physical manifestation of the cosmological scale, such as the black hole, it follows that the quantum properties of the vacuum must influence it, in significant way, on microscopic scale. Since in the realization of a black hole on smaller and smaller scale leads to mass concentration on shorter and shorter distance, we can argue that the quantum potential (positive) energy associated to the black hole(BH) increases. This fact can become important when the quantum potential energy becomes in modulus of order of the BH gravitational energy.

In this section we evaluate the lower limit of the energetic stability of a black hole taking into account the QP energy.

In order to evaluate the minimum positive value of the QP energy of a black hole we assume that the particle mass distribution inside the horizon of event approximately reads

$$| \cdot | \approx \cdot_0 \, cos \frac{2®}{4R_g} q_x \, cos \frac{2®}{4R_g} q_y \, cos \frac{2®}{4R_g} q_z \qquad - R_g < q_i < R_g \tag{32}$$

$$\cdot = 0 \qquad | q_i | \geq R_g \, ,$$

where the length a is chosen as large as possible and set equal to the BH gravitational radius (this since PDs with lower a have higher QP values and own to higher energy eigenstates [19] ) to obtain

$$V_{qu} = \frac{3\hbar^2}{m} \left( \frac{®}{2R_g} \right)^2 \tag{33}$$

where it has been assumed that that $\frac{a}{2}$ equals the diameter of the BH $2R_g$ .

By using the gravitational radius expression $R_g = \frac{2Gm}{c^2}$ united to the Planck mass $m_p = \sqrt{\frac{\hbar c}{G}}$ , it follows that

$$R_g = \frac{2Gm}{c^2} = \frac{2Gm_p}{c^2} \frac{m}{m_p} = \frac{2®\hbar}{m_p c} \frac{m}{m_p} \tag{34}$$

where $G$ is the gravitational constant, and

$$V_{qu} = \frac{3}{2} \left( \frac{m_p}{2m} \right)^3 m_p c^2 \tag{35}$$

Moreover, being the gravitational energy of the black hole $E_{grav} = -\frac{Gm}{R_s} = -\frac{1}{2} mc^2$ [28], it follows that the "binding" BH energy $E_b$ (i.e., the BH energy less the proper rest mass energy) reads

$$E_b = E_{grav} + V_{qu} = -\frac{1}{2}mc^2 + \frac{3}{2}\left(\frac{m_p}{2m}\right)^3 m_p c^2 \qquad (36)$$

The expression above. shows that the BH is stable (i.e., $E_b < 0$ ) if

$$m > \sqrt[4]{\frac{3}{2}}\frac{m_p}{\sqrt{2}} \cong 1{,}11\frac{m_p}{\sqrt{2}}. \qquad (37)$$

It must be noted that for $m > m_p$ it follows that $|E_{grav}| > |\frac{8}{3}V_{qu}|$ and that the quantum effect becomes small compared to the gravitational one.

Moreover, it must be noted that the introduction of the quantum potential quantizes the equation of motion [29] and hence the exact value of the black hole mass depends by the mass density configuration of the fundamental quantum state of the black hole.

## 5. Discussion and conclusion

The expression (16) shows that vacuum oscillations of the particle density (wave function modulus) on shorter and shorter distance are progressively less probable (i.e., suppressed). This physical effect due to the quantum potential (that confers to the particle density function a membrane-like elastic behavior, very rigid against short range variation) realizes the deterministic limit of the quantum mechanics on microscopic scale even in presence of noise [26].

Due to this quantum effect, the scale invariance of vacuum is broken and, hence, the vacuum thermal noise background of particle with mass is not white (as classically expected) but has a finite correlation length due to the quantum properties of vacuum on microscopic scale.

This physical behaviour is not a pure theoretical effect, but has a real effect that can be directly experimented by the electromagnetic black body radiation that also obeys to the quantum law. The present derivation of Planks law shows its origin from the quantum property of vacuum. It is worth noting that such characteristics of the vacuum break the scale invariance of classical physics leading to the emergence of the quantum mechanics on microscopic scale.

This fact puts under the light the fact that the cosmological mechanics cannot replicate itself on microscopic dimensions. Among the phenomena, dealing with this eventuality, we can consider the atomic-size black holes. In this case the increase of energy due to the quantum potential as a consequence of the quantum characteristics of the vacuum will hinter the low mass black hole formation leading to the stable small scale universe.

Moreover, the vacuum thermal noise background of massive particle density can also bring to the correct stochastic generalization of the hydrodynamic representation of quantum mechanics, a theory able to comprehend the classical mechanics as the large-scale limit [26].